# Pulse force nanolithography on hard surfaces using atomic force microscopy with a sharp single-crystal diamond tip.

**Alexei Temiryazev**

Kotel'nikov Institute of Radioengineering and Electronics of RAS, Fryazino Branch, Fryazino, 141190 Russia

E-mail: temiryazev@gmail.com

**Abstract**. AFM-based technique of nanolithography is proposed. The method enables rapid point by point indentation with a sharp tip. When used in tandem with single-crystal diamond tips, this technique allows the creation of high aspect ratio grooves in hard materials, such as silicon or metals. Examples of fabricated groove arrays on Si surface with 30-100 nm pitches and 5-32 nm depths are presented. Cutting of a 63nm thick metal magnetic film demonstrated. The resulting structure is studied by use of magnetic force microscopy.

## 1. Introduction.

Nanolithography based on atomic force microscopy (AFM) includes several different techniques for pattering at the nanometer scale [1,2]. One of the most common is mechanical removal of the material using the AFM tip. There are three main types of the mechanical pattering: indentation, dynamic plowing and static plowing (scratching). When indentation lithography [3-5] is performed, the tip is fixed at the certain point of the surface, and then it is pressed into the surface. Next, the tip is pulled out, moved to the next point and the process is repeated. In order to fabricate a continuous, smooth-sided groove, the distance between the points must be rather small. In plowing operation, the tip is moving parallel to the surface. In dynamic plowing [6-8], the cantilever operates in AC mode, with the tip oscillating perpendicular to the surface at all times. In static plowing or scratching [9, 10], the tip is moved in contact mode, without any perpendicular modulation. Both indentation and dynamic plowing are mainly used for nanopatterning of soft surfaces. For the hard materials scratching is the most common technique. The main tool for scratching is a stainless cantilever with a diamond pyramidal tip having a relatively large angle between the faces and a tip radius of curvature greater than 30-50 nm [11-13]. This allows a large normal force to be applied during the scratching (up to several hundred μN), without destroying the probe. Less common tools are all-diamond probes [14], ultrananocrystalline diamond probes [15], diamond tips on sapphire cantilevers [16], diamond-coated tips [17-19] and diamond-like-carbon-coated probes [20]. The performance of different probes can be compared by evaluating the results obtained in grooving some standard material, such as silicon. Several groups studied the dependence of the width and depth of the scratched grooves in silicon on the value of the load force, the speed of the scratching and the number of repetitions of the operation; diamond-tips [21], diamond-coated tips [17-19] and diamond-like-carbon-coated tips [20] were used. Note that, in all cases, the depth of grooves was substantially less than its width. Aspect ratio (depth-to-width) $R$ was about 0.1 - 0.3. In this paper, we demonstrate a method to significantly improve this parameter. The method is based on the use of silicon cantilevers with single crystal diamond tips [22]. These tips have a high aspect ratio and small (< 10-15 nm) radius of curvature of the tip apex – figure 1 and provide excellent AFM imaging. We want to draw attention to the possibility of using them for nanolithography. Since scratching is not effective with these probes, a new technique has been developed and is presented here. Since the tip is very sharp and thin, when a force is applied along the axis of the tip, it will indent any material having a lower hardness than diamond.

However, moving the needle along the surface while under load may generate a component of force perpendicular to the tip axis sufficient to break the tip. Therefore, we conducted nanolithography using the method of indentation. Note that the term "indentation lithography" has been used in [23] for scratching (static plowing) made with a nanoindentation system, not AFM. We, as previously mentioned, will describe a process of nanolithography with indentation in each point along a line. Although a drawback of this method can result in a large indentation time required for nanolithography [3], we propose to speed the process by the application of short pulses of force; we call this mode pulse force nanolithography (PFNL). The effectiveness of the proposed method has been demonstrated by creating gratings on a silicon surface and by the cutting of a strip of magnetic metal. In the latter case, the resulting structure has been studied by magnetic force microscopy and electrostatic force microscopy

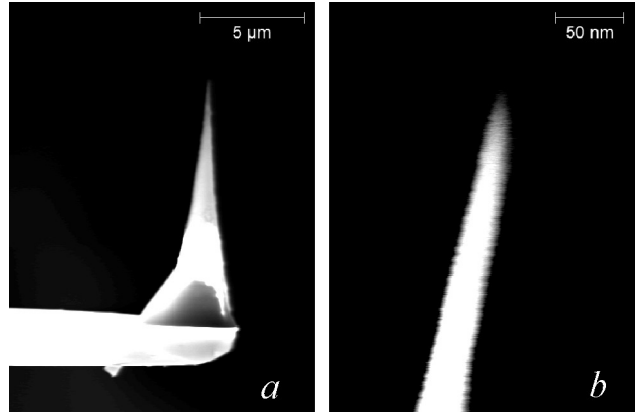

**Figure1.** SEM images of a cantilever with a single-crystal diamond tip (a) and the tip end (b)

**2. Experimental technique.**
The experiments were carried out in ambient conditions using a commercial atomic force microscope (SmartSPM, AIST-NT). The tip position in this device is fixed, AFM imaging is performed by moving the sample with capacitive sensors providing nanopositioning of the sample. When AFM scanning is performed in AC mode, the Z-feedback loop maintains a constant amplitude of the oscillating probe which ensures constant average tip-surface distance. For operation in contact mode, the Z-feedback loop maintains a constant value of the load force. The Z-piezoscanner has an unloaded resonant frequency of 15 KHz that provides high-speed Z-movement. Commercially available cantilevers with single crystal diamond tips (D300, SCDprobes) were used both for nanolithography and imaging. Typical parameters of the cantilevers are: resonant frequency 300 KHz, spring constant $k = 40$ N/m, tip aspect ratio 5:1, tip apex curvature 10 nm.

Nanoindentation is a technique usually used to measure the mechanical properties of materials. A standard test consists of taking a load-displacement curve, which plots the Z displacement of the indenter against the applied force. In routine use, a conventional nanoindenter may perform approximately one hundred indentations per day [23]. For nanolithography we need to make this process much faster, say, several hundred indentations per second. The proposed PFNL technique includes the following steps. (1) The sample moves to a point of indentation. (2) Z-feedback is disabled and a voltage pulse of duration $\tau_1$ ~ 1 ms is applied to the Z-scanner. The amplitude of the pulse corresponds to a sample displacement, $\Delta h$, towards the probe of between tens to hundreds of nanometers. (3) After a delay of $\tau_2$ ~ 1 ms after the end of the first pulse, Z-feedback is re-enabled and the sample moves to the next point with velocity $v$ ~ 1 µm/s. A distance between the neighboring points is about 5-10 nm. For soft surfaces, movement occurs in AC mode, for the hard surfaces either AC or contact mode with a small load force $f$ ~ 80 nN can be used. There are some other parameters peculiar to the PFNL technique. It is customary to indicate the value of the force applied to the tip during indentation. In order to accelerate the process of lithography we have not attempted to set a particular value for the force applied. Maximum force can be estimated as $F = k * (\Delta h - h_p - h_0)$, where $h_p$ is the depth of penetration, and $h_0$ - initial distance from the tip to the surface. When operating in AC mode, $h_0$ is determined by the mean position of the oscillating cantilever above the surface. In contact mode, $h_0$ is negative and numerically equal to the cantilever deflection when the tip is in contact with the surface $h_0 = -f/k$. Assuming that the movement takes place in the contact mode at low load force $f = 80$ nN, then $|h_0|$ is about 2 nm, and the relation $|h_0| \ll \Delta h$ is fulfilled. For hard surfaces, it can be assumed that the depth of indentation is relatively small, $h_p \ll \Delta h$, then $F \approx k * \Delta h$. This estimation will be used in discussing the results obtained.

## 3. Experimental results

### 3.1. Line arrays on a silicon surface

Line arrays were created on silicon without removing the native oxide. Between the neighboring indentation points, the sample moves in contact mode ($f$ = 80 nN) along the axis of the cantilever in the direction from the tip to the base of the cantilever. The distance between adjacent points of indentation was 5 nm. Between the lines, movement was in AC mode. For the grating with 30 nm pitch (figure 2a), the displacement of the sample during indentation $\Delta h$ was 70 nm, which corresponds to $F$ = 2.8 µN. Line arrays with 100 nm pitch were fabricated at the same displacement $\Delta h$ = 200 nm ($F$ = 8 µN) either in a single pass (figure 2b), or in three passes (figure 2c). AFM imaging of the gratings obtained was performed with the same tip in contact mode, load force $f$ = 80 nN. As shown from figure 2, the grooves are quite uniform in width and depth. Figure 3 shows the profile of grooves, two randomly-chosen cross-sections for each grating are presented. We can estimate the aspect ratio, the ratio $R$ of depth $D$ (the distance between the original surface and the lowest point in the groove) to the width $W$ (full width at half maximum). For the grating with a period of 30 nm we get mean values: depth $D = 5.1$ nm, width $W = 6.8$ nm, ratio $R = 0.75$. For the grating with a period of 100 nm, made in a single pass these parameters are: $D = 18.1$ nm, $W = 10.6$ nm, $R = 1.7$. After increasing the number of passes to three, we have $D = 31.8$ nm, $W = 13.9$ nm, $R = 2.3$. Thus the PFNL technique, in combination with sharp diamond tips, provides deep nanopatterning with high resolution. Aspect ratios of the grooves are several times larger, when compared with the results of scratching [17-21]. In our case, the time required to create a groove length of 1 μm was 1.4 seconds, with 1 second spent moving the scanner along the sample plane, and 400 ms spent for indentations at 200 points, spaced 5 nm apart with times $\tau_1 = \tau_2 = 1$ ms. Further experiments have to be carried out in order to increase the lithography speed. Limits for τ1 and τ2 depend on the scanner performance, the weight of the sample, its material, the depth of indentation, and so on. Preliminary experiments have shown that $\tau_1$ and $\tau_2$ can be lowered to 200-300 μs.

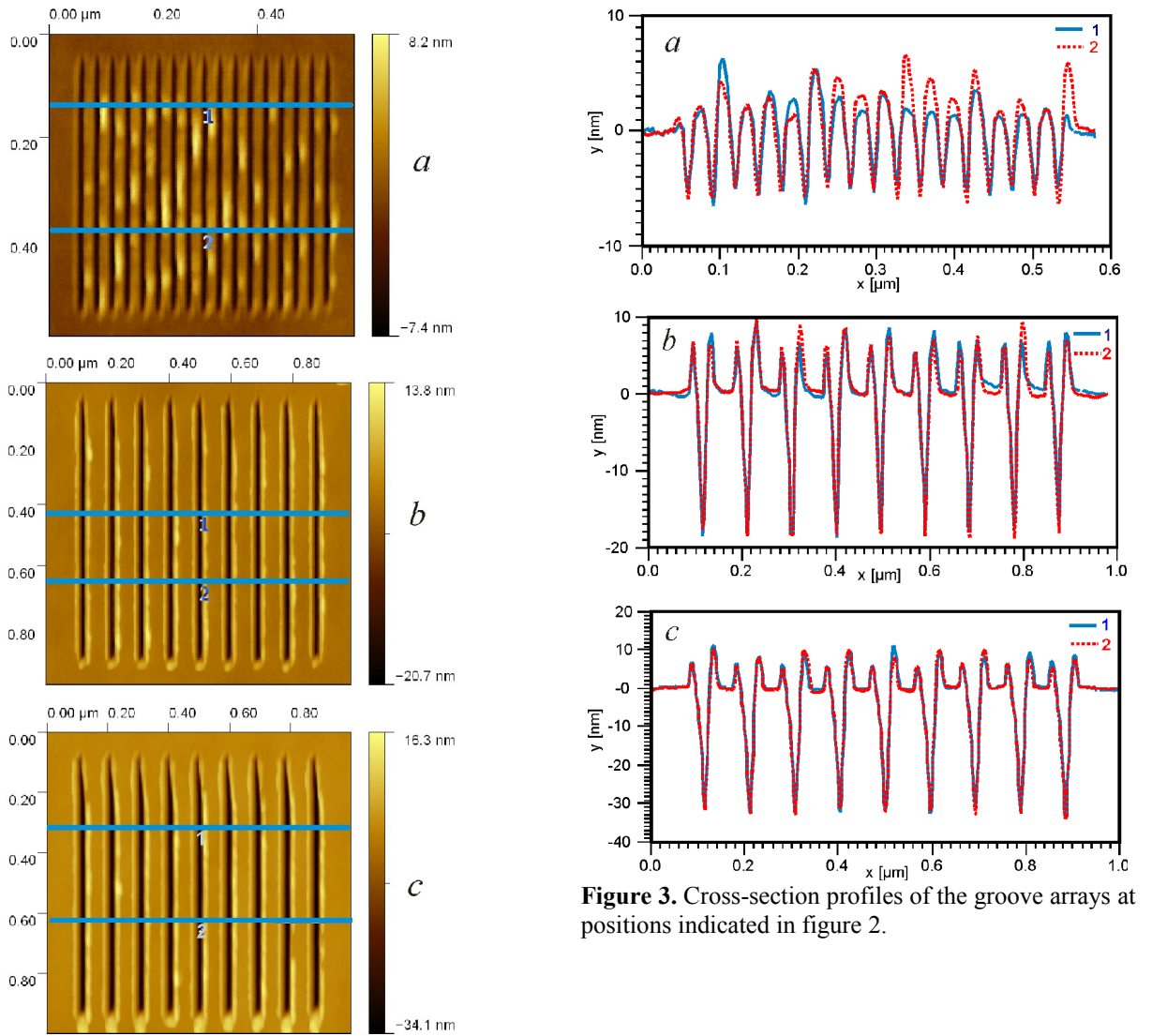

**Figure 3.** Cross-section profiles of the groove arrays at positions indicated in figure 2.

**Figure 2.** AFM images of groove arrays in a silicon surface: (a) 30 nm-pitch array made in a single pass at displacement $\Delta h$ = 70 nm; (b) 100 nm-pitch array made in a single pass at $\Delta h$ = 200 nm; (c) 100 nm-pitch array made in three passes at $\Delta h$ = 200 nm.

### 3.2. Cutting of metal magnetic films

The experiments were conducted on a 3 by 6 μm strip of CoCr film deposited on silicon oxide substrate. The thickness of the film was 63 nm. The PFNL technique was used to make two cuts, dividing the sample into three parts - figure 4. The depth of the grooves was about 72 nm, width - about 40 nm. The parameters used in cutting were: Δh = 300 nm ($F = 12$ μN), the distance between the indentation points 10 nm. The process of lithography was repeated, with an evaluation of reached depth performed during each cycle. Approximately 30 repetitions were needed until the estimated depth had clearly exceeded the thickness of the metal film. Further scanning in contact mode was performed to measure the surface profile and remove the debris created during indentation. After that, the cantilever was changed to a probe with a magnetic coating (PPP-LM-MFMR, Nanosensors) and magnetic force microscopy (MFM) [24] was used to image the domain structure of the cut sample - figure 5a. In the interpretation of the MFM image we should consider some of the results of previous MFM studies [25] of the strips, like the one that was cut. There were many strips on the substrate. The strips were placed along two perpendicular directions. Because of in-plane magnetic anisotropy, the domain structures were significantly different for the strips directed along the anisotropy axis and perpendicular to the axis. In the first case, the $180^0$-degree domain wall was directed along the length of the strip (figure 5b), in the second case the walls were across the strip (figure 5c). Initially, the element that was cut had a domain structure similar to the one shown in figure 5b. Incisions led to three structures similar to those shown in figure 5c. Thus, the PFNL technique can be used as an effective means of forming magnetic nanostructures.

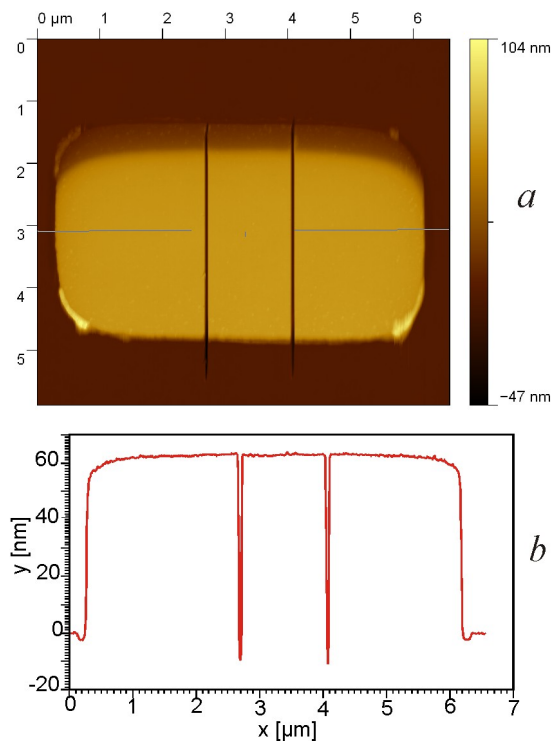

**Figure 4.** AFM image of a CoCr strip, cut into three pieces with an AFM tip (a), and a cross-section profile (b).

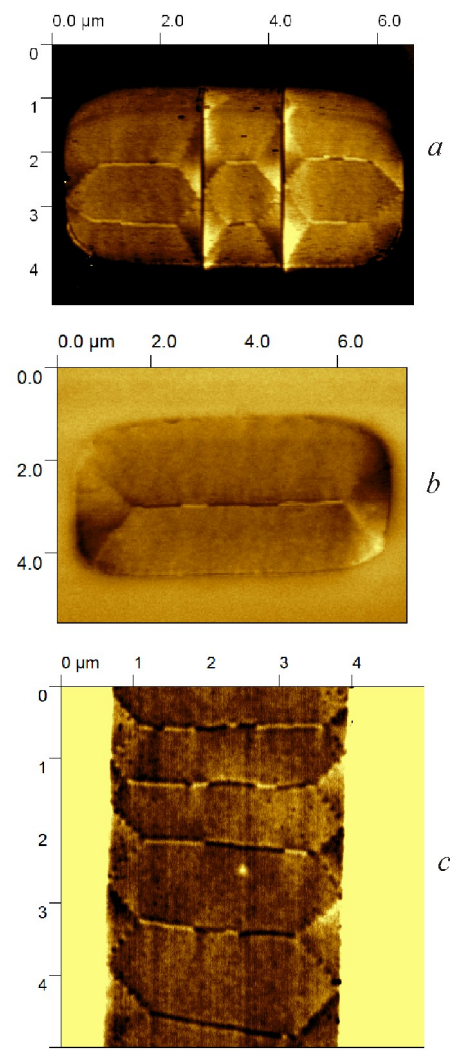

**Figure 5.** MFM images of CoCr strips: (a) the strip that is cut; (b) an un-cut strip, similar to that which was cut; (c) part of a strip lying in the perpendicular direction.

The purpose of the next experiment was to confirm that the three parts of the cut structure were electrically insulated. It is based on the idea that the electrostatic forces as well as the magnetic forces are long-range. Electrostatic forces may provide a significant contribution to the MFM image. Magnetic force microscopy makes it possible to visualize the location of domains and domain walls, the presence of Bloch lines and the structure of domain walls, but the overall contrast of a magnetic strip on the background of a non-magnetic insulator can depend strongly on the electrostatic interaction. We use this fact. Figure 6 shows two MFM images. In the first case (figure 6a), prior to scanning, a grounded tip (0 V potential) was brought into contact with each of the three parts, dissipating static charge from all parts of the sample. After that MFM imaging was performed. In the second case (figure 6b), using the same procedure the left part was charged to a potential of -7 V and the right part to a potential of +7 V, while the center part remained at zero potential. MFM image shows different contrast of sections. That means that potential equalization has not happened, the sections are well electrically insulated.

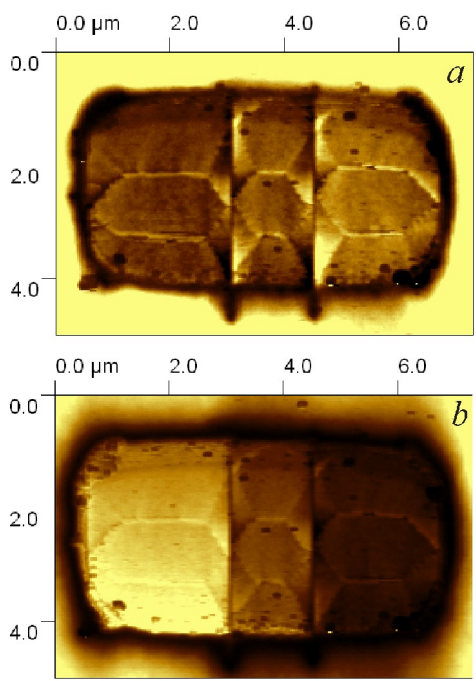

**Figure 6.** MFM images of a CoCr strip, that is cut into three pieces: (a) after removing electrical charge from all parts; (b) after charging the parts to different potentials.

## 4. Conclusions

From a general point of view, it is clear that the sharper the tool, the higher the resolution that can be obtained by mechanical lithography. In this sense, a sharp and fine diamond tip is the perfect choice. The proposed method of pulse power nanolithography allows the use of such a delicate instrument for nanopatterning of hard surfaces. The results show the promise of this technique to create nanostructures with grooves with high aspect ratios. For softer materials, standard silicon tips will be suitable.

**Acknowledgments**
The author thanks Pavel Malyshkin and Eddy Robinson for their help.